# Efficient Concept Drift Handling for Batch Android Malware Detection Models

Borja Molina-Coronado[a,*], Usue Mori[b], Alexander Mendiburu[a] and Jose Miguel-Alonso[a]

[a]*Dept. of Computer Architecture and Technology, University of the Basque Country UPV/EHU, Ps. Manuel Lardizabal 1, Donostia-San Sebastián, 20018, Gipuzkoa, Spain*
[b]*Dept. of Computer Science and Artificial Intelligence, University of the Basque Country UPV/EHU, Ps. Manuel Lardizabal 1, Donostia-San Sebastián, 20018, Gipuzkoa, Spain*

## ARTICLE INFO

## ABSTRACT

The rapidly evolving nature of Android apps poses a significant challenge to static batch machine learning algorithms employed in malware detection systems, as they quickly become obsolete. Despite this challenge, the existing literature pays limited attention to addressing this issue, with many advanced Android malware detection approaches, such as Drebin, DroidDet and MaMaDroid, relying on static models. In this work, we show how retraining techniques are able to maintain detector capabilities over time. Particularly, we analyze the effect of two aspects in the efficiency and performance of the detectors: 1) the frequency with which the models are retrained, and 2) the data used for retraining. In the first experiment, we compare periodic retraining with a more advanced concept drift detection method that triggers retraining only when necessary. In the second experiment, we analyze sampling methods to reduce the amount of data used to retrain models. Specifically, we compare fixed sized windows of recent data and state-of-the-art active learning methods that select those apps that help keep the training dataset small but diverse. Our experiments show that concept drift detection and sample selection mechanisms result in very efficient retraining strategies which can be successfully used to maintain the performance of the static Android malware state-of-the-art detectors in changing environments.

## 1. Introduction

Drift, which refers to the phenomenon where the statistical properties of the data being analyzed change over time, can be caused by data drift and/or concept drift. Data drift refers to changes which occur in the distribution of the input data over time, whereas concept drift or model drift is caused by changes in the relationship between the input data and the outcome of models, i.e., the conditional probability distribution of the class variable given the input Gama et al. [2014]. Even if both drift types are interesting and deserve analysis, it has been demonstrated that concept drift is an urgent issue in Android malware detection since it causes the trained static machine learning (ML) models to experience a steady decrease of their performance over time Pendlebury et al. [2019], Molina-Coronado et al. [2023], Chen et al. [2023a]. In this sense, in the rest of this paper, whenever we mention the term drift, we will refer to concept drift.

It is evident that the Android application ecosystem has an evolving nature, because for example, new types of malware appear or new software features are added to the development framework Molina-Coronado et al. [2023]. However, most current anti-malware research solutions for Android rely on batch ML algorithms Liu et al. [2020]. Under laboratory conditions, these algorithms have demonstrated extraordinary malware detection rates with low numbers of false positives, which make them a promising solution against malware Ucci et al. [2019]. However, batch ML algorithms are designed for static environments. They are used to train models offline on large datasets of labeled samples of malicious and benign apps, which are then used to enable accurate detection of new, previously unseen malware. Therefore, detectors that rely on these algorithms quickly become obsolete and lose effectiveness due to concept drift Gama et al. [2014], Bayram et al. [2022].

In recent years, concept drift management methods have emerged as a promising solution to the challenges posed by drift in non-stationary applications Lu et al. [2019] and in a variety of domains, including fault diagnosis Žliobaitė

*Corresponding author
✉ borja.molina@ehu.eus (B. Molina-Coronado); usue.mori@ehu.eus (U. Mori); alexander.mendiburu@ehu.eus (A. Mendiburu); jose.miguel@ehu.eus (J. Miguel-Alonso)
ORCID(s): 0000-0001-9372-5219 (B. Molina-Coronado); 0000-0002-2057-1770 (U. Mori); 0000-0002-7271-1931 (A. Mendiburu); 0000-0003-4616-322X (J. Miguel-Alonso)

et al. [2016], credit card fraud detection Blázquez-García et al. [2021], network intrusion detection Molina-Coronado et al. [2020], and game recommender systems Al-Ghossein et al. [2021]. Concept drift management methods can be classified into two major groups: (1) retraining, which consists of replacing old models with new ones trained on the latest available data, and (2) incremental algorithms, which continuously update models as new data arrives. While incremental solutions are specific learning algorithms, retraining offers the advantage of being an agnostic approach that can be applied to any ML-based detector.

For Android malware detection, several researchers have proposed adaptive solutions to overcome the challenges posed by concept drift, either relying on incremental algorithms Narayanan et al. [2017], Xu et al. [2019] or retraining procedures Karbab and Debbabi [2021], Guerra-Manzanares and Bahsi [2022]. These algorithms propose completely novel detection approaches and ignore the relevance of most available state-of-the-art Android malware detectors, which rely on static analysis of code to extract the features that represent the apps, and leverage batch ML algorithms to perform detection Liu et al. [2020]. At this point, it remains interesting whether these existing static detectors can be enhanced and adapted to changing scenarios using simple retraining mechanisms, avoiding the need to develop new detectors.

The successful implementation of retraining on existing detectors hinges upon a series of critical implementation decisions. These decisions involve establishing an retraining policy that determines when and with what data to perform the model retraining and replacement operations Webb et al. [2016]. An inadequate retraining policy may result in unnecessary, too frequent, or insufficient retraining operations that render the model unable to adapt to changes in the distribution of data Baena-Garcıa et al. [2006], Bifet and Gavalda [2007]. Equally crucial is the selection of representative data reflecting current trends in the distribution but without forgetting reoccurring or stable patterns. New data has to be continuously stored, analyzed (sometimes manually) and labeled prior to being used for retraining Android malware detectors. Moreover, as the volume of the new incoming data increases, the storage, labeling efforts and computing requirements for retraining also increase proportionally Tam et al. [2017].

The purpose of this paper is to investigate the potential of retraining as a valid approach to enhance state-of-the-art batch Android malware detectors. Indeed we focus on retraining existing detectors and analyze techniques that reduce the cost of retraining. Particularly, we focus on two critical aspects: (1) the frequency of retraining and (2) the data used for this operation. Since the factors that cause drifts and thus, model aging, could be diverse and variable, model performance is monitored to trigger an update procedure whenever a degradation of the performance is observed. Regarding the training set used for model updates, we propose strategies to keep its size small and reduce the cost of labeling new data. Thus, minimizing the cost of retraining supervised models. Through a comprehensive set of experiments, we demonstrate that retraining offers a practical solution to address concept drift in solutions that use batch ML algorithms for Android malware detection.

The rest of this paper is organized as follows. Section 2 analyzes the literature related to the present work. Section 3 introduces batch Android malware detection and how retraining can easily be applied to achieve model evolution. Then, in the next two sections, we focus on the specific methods that we analyze in this paper to determine the retraining frequency (Section 4) and the data used for retraining (Section 5). Section 6 presents our experimental setup, introduces the three state-of-the-art batch Android malware detectors used in our experiments and describes the evaluation procedure followed for the analysis. Section 7 presents the obtained results and, finally, we discuss the main findings of our work, future research lines and conclude this paper in Section 8.

## 2. Related Work

Learning in evolving environments requires defining two main aspects: 1) the mechanism used to update the model and 2) the data used to update the model. In this section, we briefly review the related proposals in the area of Android Malware detection, considering these two axes.

### 2.1. Adaptative Malware Detectors

As mentioned, the first decisive aspect when building a classifier in environments with drift is the mechanism used for adapting the model. Indeed, among the proposed adaptable Android malware detectors, we can find incremental learning algorithms that update their models with each data point, or retraining approaches, that train new models and replace the existing ones.

In a recent work Guerra-Manzanares and Bahsi [2022] propose the use of a pool of batch RandomForest classifiers and an anomaly detection model fed with system call features. Detection is performed by majority voting the output of the models. Whenever models in the pool disagree, the anomaly detector is used to conclude the class of samples. In order to enable model adaptation, true labels are assumed to be known and the worst performing RandomForest model from the pool and the anomaly detector are retrained at fixed time chunks. In Narayanan et al. [2017], the use of an incremental learning detector that leverages contextual API call information as the feature set is proposed. The model is updated with every incoming sample. However, it assumes that the true label of every sample is known at real-time. The detector proposed in Karbab and Debbabi [2021], uses a pool of Convolutional Neural Networks (CNN) fed with sequences of method, object and field names invoked in the code. Retraining is performed at fixed time chunks and using only samples for which the predictions are sufficiently reliable, so that labels obtained with majority voting of the pool are assumed to be accurate. In each retraining round the entire pool of CNN models is replaced. In DroidEvolver Xu et al. [2019], a pool of incremental linear models is presented. For updates, models with low agreement decisions with respect to the rest of the models in the pool are adapted. For labeling the data, the approach uses pseudo labels obtained through majority voting of model decisions.

As mentioned in the introduction, all these approaches are completely novel detectors, which do not leverage any previously published state-of-the-art batch detector, at least not directly. In this sense, the difference between these works and our proposal is that we attempt to directly use the existing research using model agnostic retraining policies to enhance or maintain their performances when concept drift is present. Additionally, these proposals present issues related to the labeling of samples. For instance, using pseudo labels computed from model decisions has been shown to cause model contamination over time Kan et al. [2021], while obtaining true labels incurs a cost that is often overlooked.

## 2.2. Out-of-Distribution Samples

The second aspect that must be taken into account when using retraining policies in drifting environments is the selection of data used for retraining. This data must be representative of the current concept, but the cost of labeling this data and retraining the model is proportional to the amount of data we use in this process. In this sense, some data selection strategies have been proposed in the Android literature.

The most common approach is to use the confidence of the current model in the prediction of a new sample as a way to analyze whether this new sample has been generated by the same probability distribution or not Yang et al. [2021a]. Confidence of a new sample can be measured by analyzing the consensus of several classifiers when predicting its class. In Xu et al. [2019] and Zhang et al. [2020], low confident samples (for which models disagree the most) are used to update the models. Contrary to these approaches and despite it being potentially detrimental to the adaptation ability of models, in Narayanan et al. [2017] and Karbab and Debbabi [2021] low-confident data is treated as noisy and discarded from the update process to avoid model contamination when using pseudo labels. Similarly, Barbero et al. [2022] presents a decision rejection framework which aims to keep model decisions accurate over time by discarding unreliable model decisions for drifting samples. The framework presents a non-conformity measure which identifies drifting samples with respect to a set of reference samples used to train the model.

Other authors have proposed using specific models based on clustering ideas. Yang et al. [2021b] uses a neural network based on contrastive learning to group samples into either goodware or a specific malware family. A sample is identified as drifting if it lies far from all the identified groups in a certain retraining step. This proposal has been recently improved in Chen et al. [2023b] using a hierarchical contrastive learning classifier that ranks samples according to the fitness of the CL embedding and the prediction score of the classifier. The aim is to provide a more robust drifting sample selection in unbalanced scenarios.

All these OOD (out-of-distribution) selection proposals, focus on identifying the best samples to increase the detection ability of models. However, none of them can be directly used in a simple retraining framework that is model agnostic (i.e., is built over any detector). Additionally, they are general approaches that do not leverage the particular behavior of the Android environment to design specific sample selection strategies. In this paper we will analyze, CL approaches and uncertainty sampling as model-agnostic retraining policies, and an ad-hoc sample selection method specifically designed for this problem.

## 3. Preliminary Concepts

We have discussed how most of the published literature on Android malware detection ignored concept drift as a foundational feature of Android malware detection. This section briefly describes how malware detection is typically performed using batch ML algorithms, as well as how these state-of-the-art detectors can be integrated into a retraining pipeline.

### 3.1. Batch Malware Detection

Typically, the Android malware detection process using batch ML consists of three main phases: a preprocessing stage, a training phase and a prediction phase Ucci et al. [2019]. This process is depicted in Figure 1. To begin with (preliminary step), a set of apps is required, and two tasks must be carried out. First, all the apps must be labeled. The labeling process consists of analyzing the code, metadata, and application behavior to identify any suspicious activity or known malware signatures, tagging the applications in the dataset as goodware or malware. Additionally, in this preprocessing stage, apps are examined, extracting the features indicative of their functionality and representing them in a structured manner. Examples of these features include permissions, function names, strings in the code, etc. Once the app labels and their features are obtained, in the training phase, ML algorithms help determine the most characteristic patterns of goodware and malware. As a result of this training stage, a ML model capable of predicting the class label (goodware or malware) of new apps is obtained. Finally, the prediction phase consists of extracting the features identified during the training phase from a new incoming app. Afterwards, these features are fed into the previously trained ML model so that it determines whether the app is goodware or malware.

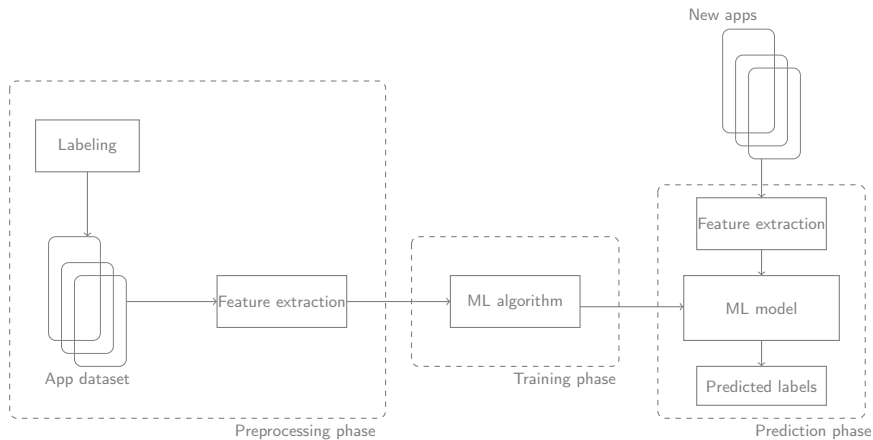

**Figure 1:** Diagram of the batch learning process. (1) Preprocessing phase: a structured feature set and a label is obtained for each app; (2) Training phase: the structured and labeled training dataset is used to generate a model using ML algorithms; (3) Prediction phase: the generated model is used during the prediction phase to determine the class of new apps.

### 3.2. Retraining for Batch Malware Detectors

Retraining mechanisms consider a detector as a black box tool. This means that any existing batch detector can be integrated into the retraining process without modification. Figure 2 depicts how the retraining mechanism can be integrated into any existing detector. In order for new models to correctly represent the current data distribution, the training data has to be continuously updated with representative apps. Since Android malware detectors rely on supervised algorithms, these apps must be labeled. Retraining is signaled by a supervisor. Whenever the signal is raised, a new model is trained to replace the old one. This involves preprocessing all (or some) apps in the dataset to extract their features, training the new model with this information, and replacing the old model.

A very simple retraining policy is to activate the update process at fixed time intervals, for example, once a month. It can also be triggered whenever a certain number of new labeled apps become available, e.g., when 10 000 new apps have been identified. Nonetheless, the most effective strategy would be to trigger retraining whenever a drift is detected. The supervisor can monitor the performance of the model, or measure the degree of dissimilarity between training apps and incoming apps. In the following sections, we investigate the impact of some of these retraining strategies, as well as the impact of different retraining data management policies on the efficiency of batch Android malware detectors. Particularly, we focus on two mechanisms: fixed-period retraining and using a monitor that identifies changes to trigger updates. In addition, we explore three approaches for managing retraining data: a forgetting mechanism that discards old apps, three active learning methods that select highly-relevant data and a sample selection technique that removes uninformative data.

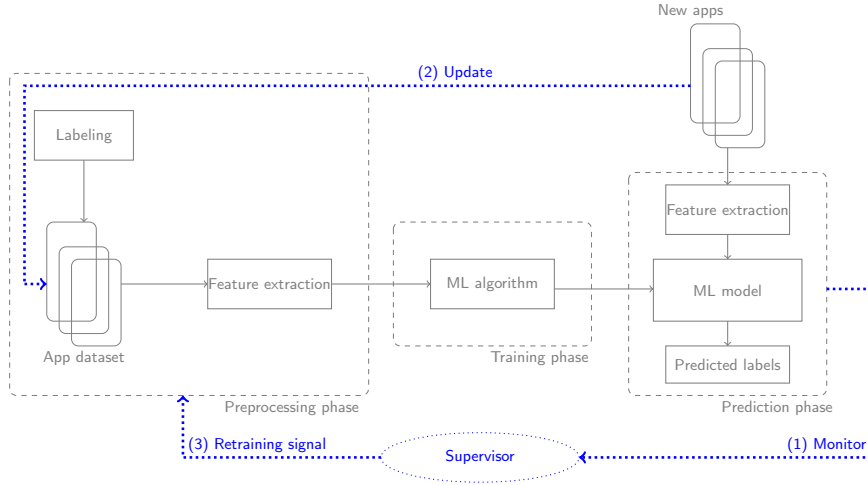

**Figure 2:** Diagram of the batch learning process with retraining. The supervisor firstly monitors when changes take place, once a change is detected, the data is updated to reflect the current trend and a model retraining signal is raised. This trains a new model with the updated data that is used to replace the old model.

## 4. Retraining Frequency

In this section, we discuss the two different retraining policies mentioned above: (1) scheduling the update operation periodically and (2) using a change monitor that triggers the update when the performance of the detector drops.

### 4.1. Fixed Period Retraining

A naive update policy is to retrain Android malware detectors in batches using a fixed periodicity: weekly, monthly or any other. This method has several advantages, including ease of implementation and predictability. By following a fixed schedule, the system can regularly retrain the model to keep it up-to-date with the latest malware trends and behavioural patterns. However, this approach also has some limitations, because the rate of change of the data distribution might not be uniform or periodical. Due to the unpredictability of changes in Android data, choosing a fixed update frequency may be suboptimal. If the time between updates is long, the model may miss malware that has appeared and lasted for a short period of time. On the other hand, a high retraining frequency would eliminate this problem, but would result in unnecessary costs if changes in the data distribution are slow Gama et al. [2004].

### 4.2. Change Detection Mechanisms

An alternative update policy to fixed period retraining is to use a change detection mechanism which monitors the current data or the performance of the model, triggering an update round only when there is evidence of change.

For this purpose, in this paper we consider the Page-Hinkley (PH) test Page [1954], Hinkley [1970], a popular (and easy to implement) drift detection algorithm that detects changes by monitoring the performance of the model. The PH test has several advantages over other change detection methods. First, it is non-parametric and does not make any assumptions about the underlying data distribution. Secondly, it is computationally efficient and requires minimal memory, which makes it suitable for monitoring high-speed data streams. Finally, it is also robust to outliers and can detect gradual changes in the data distribution Bifet and Gavalda [2007].

$$C_n = \begin{cases} 0 & \text{if } n = 1 \\ \min\left(0, C_{n-1} + (A_{mean_n} - \bar{x}_{n-1})\right) & \text{if } n > 1 \end{cases} \quad (1)$$

$$\bar{x}_{n-1} = \frac{\sum_{t=1}^{n-1} A_{mean_t}}{n-1} \quad (2)$$

| $TPR = \frac{TP}{P}$ | $TNR = \frac{TN}{N}$ | $A_{mean} = \frac{TPR+TNR}{2}$ |

**Table 1**
Metrics used in this paper to assess the performance of detectors. TPR = True Positive Rate. TNR = True Negative Rate.

$$PH_n = \begin{cases} 1 & \text{if } \lambda + C_n < 0 \\ 0 & \text{if } \lambda + C_n >= 0 \end{cases} \quad (3)$$

The PH test is applied as follows: it periodically (or whenever a certain batch of new instances are obtained) monitors a test value calculated based on the performance of the model, in our case, measured by the $A_{mean}$ (see Table 1). Specifically, at each instant $n$, the PH method computes the CUSUM ($C_n$) of the deviations between the current performance value ($A_{mean_n}$) and the mean of the performance values obtained in all the previous periodic checks (see Equations 1 and 2). If the CUSUM of the deviations falls below a pre-defined $\lambda$ threshold (see Equation 3), the PH test signals a change ($PH_n = 1$) that triggers a model update at instant $n$. Note that a higher tolerance value may result in a lower rate of false alarms, but also in a lower performance, as updates can be delayed. When a change is detected, the values used for the test are reset. This means that the instant $n$, at which the test flags the change, is set as the starting point (0) for subsequent calculations of the test.

## 5. Data Used for Retraining

The effectiveness and efficiency of retraining also depends on the data used to update the models. In this section, we analyze the use of fixed-size sliding windows and active learning methods such as uncertainty sampling, contrastive learning OOD methods and a problem-specific sample selection strategy.

### 5.1. Sliding Windows

In this approach, a fixed-size sliding window is used to select the $m$ most recent instances for retraining. The window moves forward whenever new data becomes available, and instances that fall within the window are stored and subsequently used for retraining, while older apps are discarded. We depict this policy in Figure 3. Its implementation is straightforward, but it may have some drawbacks. First, it assumes that the $m$ most recent apps are sufficiently representative to generate a good model, which may not always be true: behaviours of discarded apps may reappear later. Secondly, this method does not consider the characteristics of the instances within the window. A common feature of the Android app environment is the presence of majority groups, that is, apps that are nearly identical and appear in large quantities. Not considering the characteristics of the last $m$ apps might result in datasets where some types of apps are over-represented while others are largely underrepresented, thus leading to biased models which ignore the minority samples Gonçalves Jr et al. [2014].

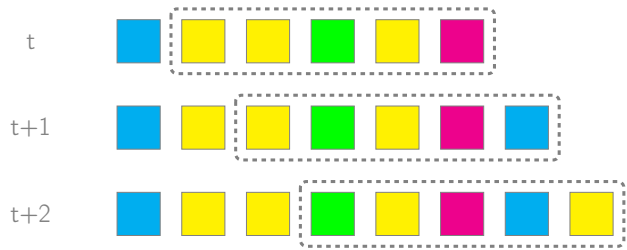

**Figure 3:** Sliding window (dotted line) of size $m = 5$ which is used to train the model at each instant. Colored cubes represent apps with different behavioral patterns and whose predominance may vary over time. A model trained at time $t$ may be biased towards the "yellow" behaviour, while being unable to recognize the "blue" behavior.

## 5.2. Uncertainty Sampling

Uncertainty sampling is a technique commonly proposed in the active learning literature to reduce labeling efforts and improve the learning ability of models Yang et al. [2021a]. The method measures the reliability or uncertainty of the decisions provided by models for samples. The degree of uncertainty of a sample is computed as the complementary of the absolute difference of the class (goodware and malware) probabilities returned by a model. Since low confident decisions for samples result in similar probability values for both classes, the uncertainty value will be high (close to 1), whereas samples where one class probability dominates the other will obtain low uncertainty values close to 0. The method assumes that samples with the highest uncertainty are the most representative of changes and better candidates for learning a new model. Therefore, two alternative criteria can be used to build the training dataset: (1) set a fixed number *n* of samples to select and, (2) set a minimum uncertainty value to select the samples. Finally, the selected samples are added to the samples used for the previous retraining period and a new model is built with all this data.

## 5.3. Contrastive Learning OOD

More advanced sampling mechanisms use Contrastive-Learning (CL) schemes that rely on an encoder-decoder architecture to identify drifting samples. The CL model is trained to generate similar embedded representations (or embeddings) for samples of a same class or malware family, whereas the embeddings for samples of different classes (malware vs goodware) or malware families are dissimilar. Since the CL encoder identifies the characteristics in the training data that help to separate the samples that pertain to different classes, CL sampling methods measure the dissimilarity of new samples according to how their embedding differs from those of the training data Yang et al. [2021b], Chen et al. [2023b]. Similarly to uncertainty sampling methods, a ranking is constructed based on the dissimilarity measure and samples are selected by: (1) selecting the *n* most dissimilar samples, or (2) setting a threshold over the dissimilarity measure as the minimum sample selection criterion. Afterwards, the selected samples are appended to the training samples of the previous retraining period.

## 5.4. Problem Specific Sample Selection

As mentioned previously, in the context of Android malware detection, apps with some specific features may be more prevalent than others. Recurring malware that fades away and resurfaces is also a reality. To exemplify this, Figure 4 depicts the distribution of goodware and malware into known and unknown behaviors for every quarter between January of 2013 and December of 2019. In this context, an app behavior is represented by a particular set of API call frequencies extracted from its code. We assume that apps with similar behavior present equivalent API call frequencies in their code. The exact process for the computation of known and unknown behaviors is explained more in detail later in this section. Green slashed bars represent the proportion of samples on each period that contain similar behaviors to apps observed in previous periods (known), whereas grey dotted bars represent the proportion of samples whose behavior has not been observed previously (unknown). As can be seen in Figures 4a and 4b, the apparition of unknown app behaviors from one period to another confirms the existence of data drift in the Android application ecosystem, which can cause model degradation Chen et al. [2023a]. It also shows that the incidence of drift is variable (for example, differences in malware between 2015Q2 and 2015Q3). In this regard, goodware tends to present more novel patterns over time, whereas malware frequently exhibits more known behaviors that have been observed in preceding periods. Indeed, the fluctuations observed for the malware follow a common infection pattern. Each time a new form of infection emerges, the apps (samples) exploiting this method will be initially classified as unknown (see, for example, 2015Q1). Then, the infection mechanism becomes popular as new malware apps use it. This is shown, for example, by the increase of known groups in the subsequent periods to 2015Q1. This popularity keeps increasing until the infection pattern is detected and a new form or a variation of the original exploitation mechanism is developed.

The over-representation of malware with known behaviors during most periods can lead to biased detectors when retraining ML models, as algorithms are designed to optimize performance metrics and may focus solely on these majority groups Zhao et al. [2021]. Hence, using all the data for training can hinder the ability of detectors to accurately distinguish minority (unknown or new) malware. With the aim of improving the effectiveness of the adaptation mechanism and producing more reliable malware detectors, we propose the use of an ad-hoc sample selection approach for Android malware detection. This technique ensures that the retraining data is diverse and informative Molina-Coronado et al. [2023]. It involves filtering out uninformative or duplicated apps, controlling the size of the dataset, and reducing the labeling costs and training complexity of ML algorithms.

Particularly, in this work we propose a sample selection method using the continuous clustering process described in a previous work Molina-Coronado et al. [2023] and initially proposed in Portnoy [2001]. For this algorithm, and based on previous findings, we represent the apps as a vector of frequencies of their Android API calls. Then, sample selection is carried out in two phases: the calibration phase and the online phase.

The objective of the calibration phase is to find the different behavioral groups present in the training data, for both malware and goodware. To do so, the apps in the training set are chronologically ordered by their publication date and sequentially assigned to their closest cluster. This assignment is only performed if the sample lies within a predefined $\epsilon$ radius from the cluster's representative; otherwise, a new cluster is created with the sample as the representative. We assume that samples within a group contain similar code patterns and thus, that each cluster represents a particular behavioral pattern. Note that cluster's representatives are maintained throughout the process. The Euclidean distance is used to measure the similarity between every pair of samples. Once all the apps are clustered, we label the clusters as goodware or malware according to the class label of the representative app of that cluster. Within this calibration phase we compute the average number of apps in all the behavioral clusters found ($k$). Then, we only keep the most recent $k$ components (apps) from each cluster. In this way, we try to keep the training set both small (keeping only a few samples of a given behavior) and diverse (keeping samples of all the different behaviors detected).

During the online phase (concept drift handling), the algorithm assigns each new incoming sample to its closest cluster if it meets the admission condition (the sample is within the $\epsilon$ radius of the representative). If the cluster already contains $k$ samples, the oldest sample in the cluster is replaced by the new one. This process can be seen as a multi-window approach in which a sliding window of size $k$ is maintained for each of the behavioral clusters. If a sample cannot be associated with an existing cluster, a new cluster is created with that sample as its representative. At the end of the clustering process, we compute the isolation level of clusters as the average Euclidean distance between the cluster representative and the representatives of other clusters. For labeling, we select the representatives of the $l_b$ most isolated clusters, being $l_b$ a labeling budget parameter. Finally, the retraining dataset is constructed by appending to the samples used on the previous retraining period, the $k$ most recent samples of each labeled cluster. Note that the apps that are assigned to a cluster are automatically labeled with the class label of the cluster representative. This avoids labeling many apps.

## 6. Experimental Framework

This section describes the experimental set-up and the methodology followed to evaluate the different adaptation mechanisms analysed in this work.

### 6.1. Dataset

In our experiments, we use the dataset presented in Molina-Coronado et al. [2023]. This dataset consists of eight years of malware and goodware sorted in a monthly basis, from January 2012 to December 2019. In the preprocessing step, class labels are assigned based on the number of VirusTotal detections (VTD) Kantchelian et al. [2015]. Apps with a VTD value equal to 0 are tagged as goodware, and apps with a VTD value greater than or equal to 7 are tagged as malware. The remaining apps (those with a VTD value between 1 and 7) are discarded. This labelling methodology is common in the Android malware literature Zhu et al. [2020], Salem et al. [2019]. The instructions to download the dataset are available in our GitLab repository[1].

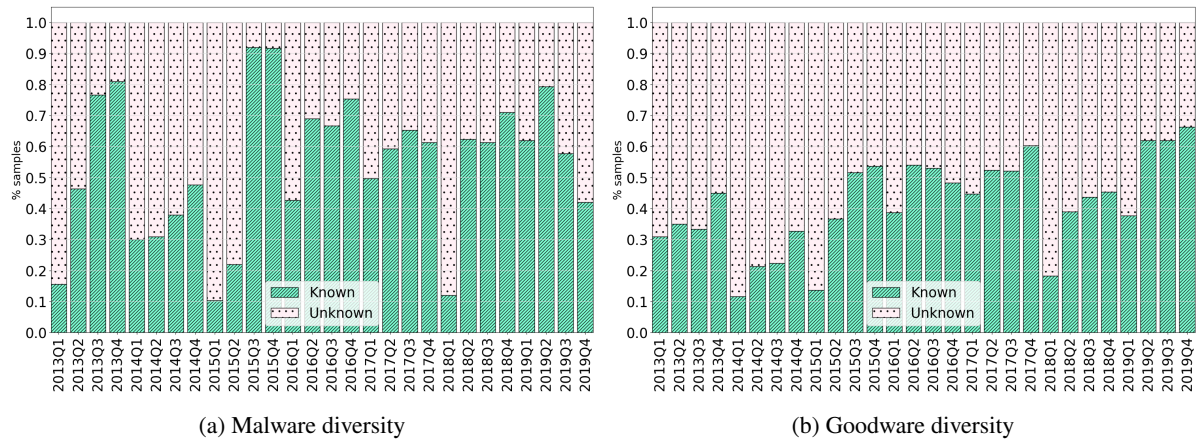

(a) Malware diversity    (b) Goodware diversity

**Figure 4:** Diversity of the dataset throughout the evaluation period (2013-2019). The bars represent the percentage of apps captured in the indicated period that are very similar to apps that have already appeared in previous periods ("Known"), or apps that exhibit new behaviors ("Unknown").

Once this is done, we split this dataset into two separate subsets: one for training and one for evaluation. The training dataset consists of 100 monthly samples of each class, goodware and malware, between January 2012 and December 2012. Note that malware detection on Android is a highly unbalanced problem where malware actually accounts for about 10% of the apps Pendlebury et al. [2019]. However, the training dataset is compiled offline and, thus, can be constructed using an unrealistically balanced ratio between the two classes. Contrarily, in order to mimic a real situation, the remaining 7 years of data (from January 2013 to December 2019), used for model evaluation purposes, will consists of 10 malware and 100 goodware samples per month over this period, which are obtained by randomly sampling apps from the original dataset.

### 6.2. Batch Malware Detectors

For the purpose of this paper, we rely on three state-of-the-art malware detectors, Drebin, DroidDet and MaMaDroid, that, according to a recent comparison Molina-Coronado et al. [2023], are the best performing batch Android malware detectors published to date. These detectors were not originally conceived to cope with concept drift and they all rely on features extracted through static analysis of APK[2] files to represent the apps, and on batch models for detection. In this section we briefly describe their detection mechanisms:

**Drebin** Arp et al. [2014] uses a full set of features extracted from APKs, including hardware components, permissions, application components, intent filters, strings, and a restricted set of API calls. It uses a linear SVM model fed with this data to perform malware detection.

**DroidDet** Zhu et al. [2018] relies on data obtained exclusively from the app code to detect malware. More specifically, it uses a filtered set of requested and required permissions, intent filters and API calls. After the first extraction of all possible values, the relevance of the features is calculated to eliminate those that are not informative. The most relevant features are finally used for model generation using the RotationForest algorithm.

**MaMaDroid** Onwuzurike et al. [2019] constructs a Markov chain of the API calls found in the app code. The Markov chain represents the transition frequency between each API pair. Actually, the package to which an API call belongs is used as a higher level abstraction to reduce the number of final features. The RandomForest algorithm is used to identify malware with this information.

### 6.3. Parameter Settings

The baseline (original) detectors have been trained using the default parameters reported in their respective works. For all configurations, the evaluation and (possible) retraining process is set to quarterly intervals. This choice is a compromise between obtaining an adequate visualization of the results but restricting the number of new models, since training the models has a significant experimental cost. In addition, for the change detection method, after preliminary experiments, we set the $\lambda$ threshold for the PH test to 0.02 based on preliminary results (see Appendix A.1), as a threshold between detection performance and the number of retraining steps.

In relation to the methods proposed for selecting the data used for retraining, sliding windows of 100, 1000, and 2000 apps are considered in the experimentation. For the problem-specific sample selection strategy, based on preliminary tests (see Appendix A.2), we set the $\epsilon$ radius (the maximum distance allowed to consider any sample as part of an existing cluster) to 0.01. The $k$ value (the average number of apps in a cluster), has been calculated in the calibration phase, taking a value of 2. For CL methods (CADE Yang et al. [2021b] and HICL Chen et al. [2023b]), we use the original implementation and the parameters that reported the best results in their respective papers. Additionally, since uncertainty and CL methods require setting a criterion for selecting the samples to be labeled and appended to the retraining dataset, either by taking the $n$ most uncertain samples or by setting a threshold over the uncertainty measure, we set a labeling budget similar to the average number of samples to be labeled with the problem-specific sample selection method.

---

[1] https://gitlab.com/serralba/concept_drift
[2] Android Application Package, i.e., the file format used by Android to distribute applications.

## 6.4. Evaluation Framework and Metrics

First, all models are trained offline (batch) using the apps in the training dataset (balanced and with data from Jan. 2012 to Dec. 2012). For evaluation purposes, we consider non-overlapping windows of three-month periods. Therefore, the evaluation dataset is divided into 28 time-ordered subsets, each one covering one quarter.

For the evaluation of the original version of the detectors (pure batch scenario without retraining), the model used is always the same, i.e., the one obtained in the offline phase. For those scenarios incorporating concept drift management approaches, model update procedures are subsequently carried out with a subset of recent apps. Note that we assume that, when a model is updated, the true labels of the samples used to train the new model are known. As the incoming data are chronologically sorted, we can evaluate the degree of concept drift, as well as the effectiveness of the measures implemented to address it. This approach is common in the concept drift literature Gama et al. [2014].

In two separate experiments we analyze and compare the effect of: (1) the policies to trigger the updates, and (2) the data used for retraining. In the first experiment, periodic vs. change detection mechanisms are compared. In this experiment, the dataset used for training the models grows in each retraining round since all incoming samples are incorporated to the dataset for retraining. In the second experiment, where the different data selection mechanisms are studied, the models are retrained at each trimester with the corresponding selection of data (windows of fixed size, uncertainty samples, OOD samples or cluster representatives). As a baseline for this second experiment, we also consider the model retrained periodically each trimester using all the data available.

Due to the large amount of data that is available for training and in order to avoid imbalance between the malware and goodware when retraining the models, for all the possible combinations and methods excepting those using CL or uncertainty sampling, in each retraining round, goodware is downsampled to reach a balanced ratio between the classes. Specifically, when the training dataset is constructed using the problem-specific sample selection method, once the clustering has been carried out and the goodware and malware samples are obtained, the goodware is downsampled to reach a balanced dataset. Note that this is only done for training, whereas for evaluation the original unbalanced data is used.

Finally, in all the experiments and for each model, we measure its performance as the average of the TPR and TNR, known as the $A_{mean}$ value (see Table 1). The $A_{mean}$ is a popular performance metric in the ML literature for unbalanced scenarios and, contrary to the F1 score, the $A_{mean}$ considers and equally weights the accuracy of models on both positive (malware) and negative (goodware) samples.

## 7. Experimental Results

This section shows the results of the different retraining configurations tested for the state-of-the-art malware detection models: Drebin, DroidDet and MaMaDroid. Code implementations for all these mechanisms are available in our GitLab repository[3].

### 7.1. Analysis of the Effect of the Retraining Frequency

The results when retraining the detectors at fixed periods and with change detection are shown in Figure 5. The lines in the figures represent the $A_{mean}$ performance of the models over the evaluation period. In particular, the red lines show the performance of the detectors when a periodic retraining approach is applied. The blue lines represent the performance of detectors implementing the change detection mechanism based on the PH test. The vertical dotted blue lines represent the points at which the PH test has triggered a drift alarm and, thus, a retraining and model replacement operation has been performed. For comparison purposes, we also include the performance of the (original) batch model which is trained only once, at the beginning. This is represented by a dashed orange line.

As can be seen, the orange lines show a decreasing trend over time for all models, confirming the existence of concept drift. The benefits of using retraining as an adaptation mechanism to counteract the effect of concept drift in batch malware detectors are readily apparent from the figures. For all adaptive solutions, the performance of the models is kept stable over time. In fact, the retraining variants of DroidDet (see Figure 5b) show an overall performance improvement with respect to the static version of 15%, while for Drebin and MaMaDroid this performance increases 23% and 16%, respectively (see Figures 5a and 5c).

Overall, when comparing the two retraining configurations, the figures indicate that applying a change detection mechanism has a minimal cost in performance ($A_{mean}$), with an average reduction of 2.3% for all detectors. Conversely, the change detection method requires a much smaller number of retraining operations compared to retraining at fixed periods. In fact, as can be seen, the change detector successfully triggers a drift alarm when the performance of the detectors decreases. For DroidDet, eight rounds of retraining and model replacement are required, as shown by the blue dotted lines in Figure 5b, which contrasts with the 28 operations performed with fixed-period retraining. With equivalent detection performance indicators, only seven drift alarms are triggered in MaMaDroid (see Figure 5c) and Drebin (see Figure 5a).

---

[3]https://gitlab.com/serralba/concept_drift

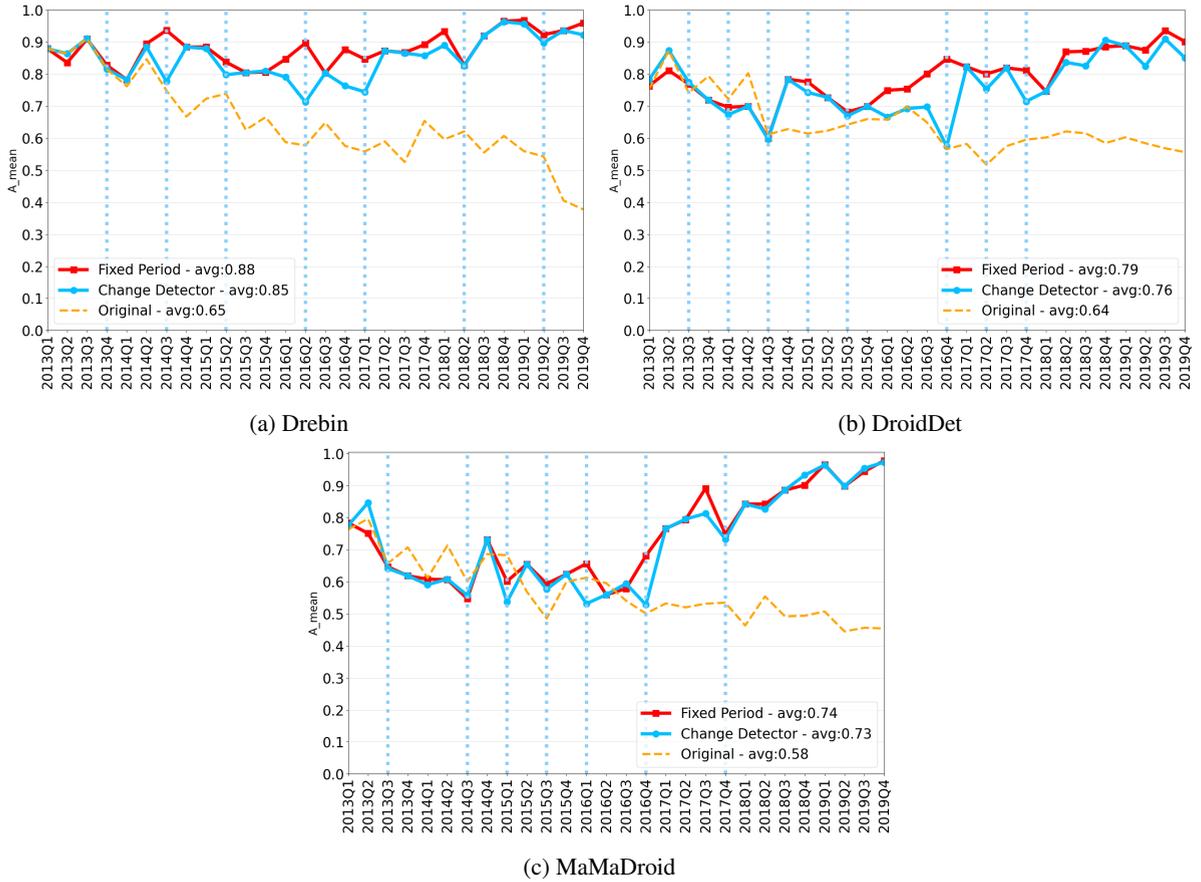

**Figure 5:** Evolution of the performance of malware detectors for the period 2013-2019, for different policies to trigger retraining. Dotted, blue vertical lines indicate a model change triggered by the concept drfit detection mechanism.

### 7.2. Analysis of the Effect of the Retraining Data

Table 2 shows the average $A_{mean}$ performance of the detectors using different data management policies when periodically re-training the models. It is notable that the use of a data management policy obtains very similar performance values or even outperforms the baseline configuration (the one using all available data) in most cases. From the tested configurations, using the problem-specific sample selection approach for retraining with a labeling budget of 70% of the incoming samples seems to be the best approach, followed by HICL and uncertainty methods with similar labeling budget. In general, except for CADE, the results do not show significant differences among active learning methods, and even using smaller datasets with 45% of labeling effort, the performance indicators remain very similar or even outperform the baseline that uses all the data.

Figure 6 shows the results for each individual detector over the entire evaluation period. For clarity of the results, we only selected the best sliding window policy, contrastive learning OOD method, and problem-specific sample selection configuration. The red lines represent the baseline, that uses all available data for retraining; the green lines represent the performance when considering a sliding window of size 1000 for retraining, the orange lines represent the performance using the HICL method with a labeling budget of 70%; and the blue lines show the performance of the problem-specific sample selection mechanism with 70% of labeling budget. Using active learning mechanisms to select samples for retraining results in improved performance values with respect to the fixed-size configuration for all methods except for DroidDet, with the problem-specific method yielding slightly better $A_{mean}$ values in most evaluation rounds than the HICL method.

|  | Drebin | DroidDet | MaMaDroid | Avg. Perf. | % Labels Req. |
|---|---|---|---|---|---|
| **All data** | 0.88 | 0.79 | 0.74 | 0.80 | 100% |
| **Last 1000** | 0.85 | 0.80 | 0.72 | 0.79 | 100% |
| **Last 2000** | 0.87 | 0.79 | 0.72 | 0.79 | 100% |
| **Last 100** | 0.75 | 0.69 | 0.68 | 0.70 | 30% |
| **Problem-Specific** | 0.88 | 0.78 | 0.82 | 0.82 | 70% |
| **CADE** Yang et al. [2021b] | 0.83 | 0.76 | 0.74 | 0.77 | 70% |
| **HICL** Chen et al. [2023b] | 0.88 | 0.78 | 0.75 | 0.80 | 70% |
| **Uncertainty** | 0.86 | 0.79 | 0.75 | 0.80 | 70% |
| **Problem-Specific** | 0.86 | 0.76 | 0.78 | 0.80 | 45% |
| **CADE** Yang et al. [2021b] | 0.77 | 0.74 | 0.72 | 0.74 | 45% |
| **HICL** Chen et al. [2023b] | 0.87 | 0.77 | 0.75 | 0.80 | 45% |
| **Uncertainty** | 0.83 | 0.80 | 0.74 | 0.79 | 45% |
| **Problem-Specific** | 0.84 | 0.73 | 0.76 | 0.78 | 15% |
| **CADE** Yang et al. [2021b] | 0.69 | 0.75 | 0.70 | 0.71 | 15% |
| **HICL** Chen et al. [2023b] | 0.86 | 0.75 | 0.74 | 0.78 | 15% |
| **Uncertainty** | 0.83 | 0.72 | 0.73 | 0.76 | 15% |

**Table 2**
Average $A_{mean}$ performance throughout the evaluation period for different sample selection policies using fixed period retraining. The right column refers to the percentage of samples in the buffer that need to be labelled for retraining.

Beyond detection performance, the effort required to label the samples used in each round of retraining is also an important factor to measure the efficiency. Considering that a total of 330 apps arrive in each retraining round (300 goodware and 30 malware), the labeling requirements for the strategies "Last 1000" and "Last 2000" are similar to those of the baseline method, as they involve labeling all new arriving samples before retraining. In contrast, the "Last 100" strategy requires labeling only about 30% of the incoming samples in each evaluation round. Active learning methods require labeling only 45% of the incoming samples to obtain equivalent performance values to the baseline and the last 1000 and 2000 sliding window policies. With a lower labeling budget, the problem-specific and HICL methods obtain very similar performance on average. These results demonstrate how detection models benefit from the use of incremental clustering to label samples and reduce the size of the training data. As a potential drawback, note that this process can lead to labelling errors. In this regard, our experiments showed that only 0.05% of the samples are mislabeled by the method, demonstrating to be insufficient to negatively impact the detection ability of ML algorithms.

For the interested reader, we also include the analysis of the combination of change detection and different sample selection methods, as well as the results obtained with different parameter configurations of the proposed methods in Appendix B.

## 8. Conclusions

In this paper, we have shown that retraining is an effective mechanism for dealing with concept drift in batch Android malware detectors, it being straightforward to incorporate into existing detectors without modifying their design. Specifically, our experiments show that this update mechanism helps maintain high detection rates, with an average performance improvement of 20% compared to the original versions of the detectors. Regarding the two retraining alternatives tested, there are no significant performance differences between periodic retraining and the PH-based change detection approach. However, using a supervision mechanism based on the PH test showed to decrease the number of retraining rounds by 75% on average, dramatically reducing the computational effort required to keep model performance over time.

Additionally, we have demonstrated that the sample selection strategy used for retraining also influences the success of detectors. On one hand, employing a sample selection policy instead of using all available data for retraining reduces the cost of model generation since the complexity of machine learning algorithms is highly dependent on the size of the training data Hastie et al. [2009]. Sliding window policies, such as selecting the last 1000 and 2000 samples, helped reduce retraining complexity while maintaining similar labeling effort and performance values compared to the baseline. The benefits of using active learning techniques, such as uncertainty sampling, HICL, or the problem-specific sample selection method, are undeniable. These techniques result in better detection performance for models and require reduced labeling effort for retraining. Among the active learning methods, the proposed problem-specific strategy exhibited minimal performance degradation under widened labeling constraints compared to other alternatives.

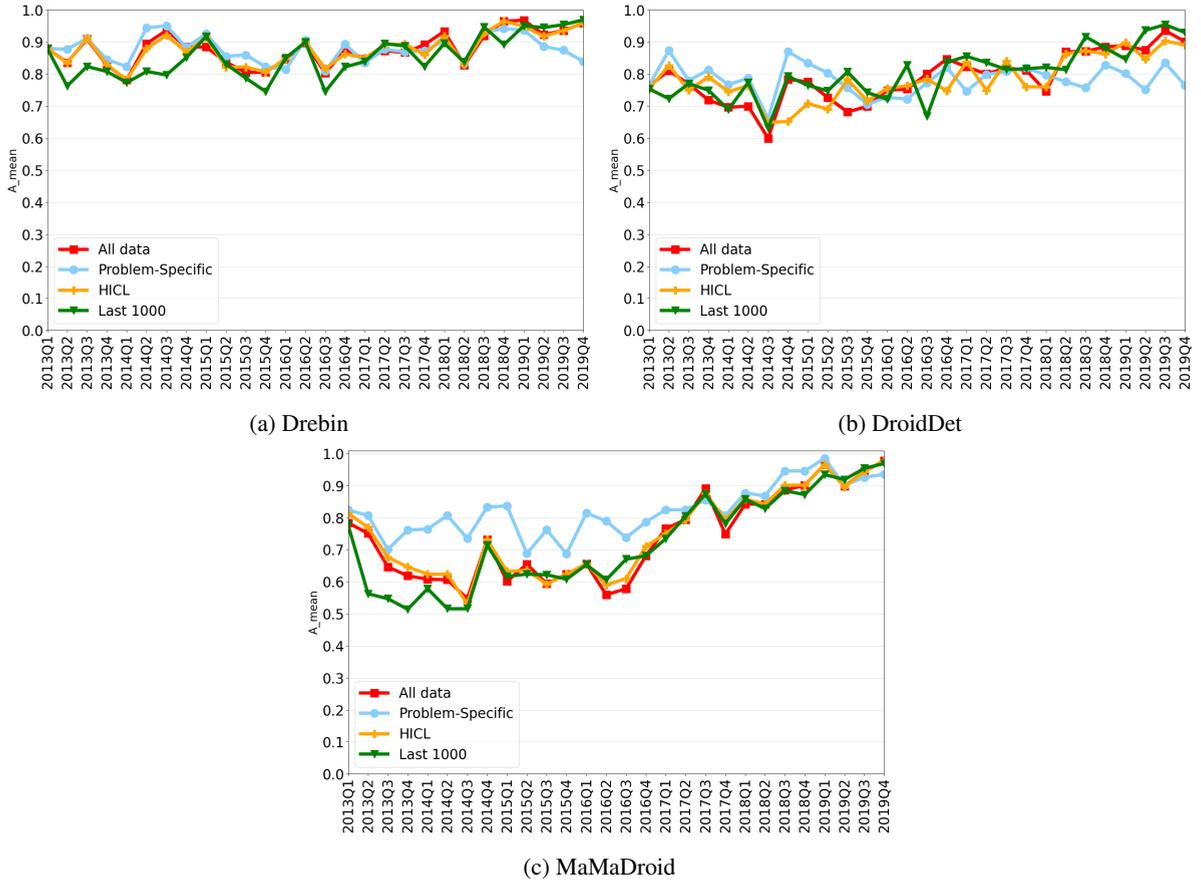

**Figure 6:** Evolution of the performance of malware detectors with periodic retraining for the period 2013-2019. Red lines represent the performance when using all the data available, i.e., the baseline. Green, blue and orange lines represent respectively the performance of models when retrained with the 1000 most recent samples, using the problem-specific strategy and with the HICL selection methods with labeling budgets of 70%.

In general, the choice of a specific sample selection strategy and retraining policy will depend on the requirements of the target scenario for the detector. Change detection is a suitable method in most scenarios, especially in cases where the cost of generating models is high. One advantage over periodic retraining is that it requires fewer retraining operations to keep models up-to-date, consequently reducing the need for labeling new samples. Labeling new samples is often costly, and in many cases, it is performed manually by human experts. In this context, the application of a sample selection mechanism is also desirable. Larger sliding window sizes need labeling all incoming data, which may not be feasible, especially in online scenarios. Shorter windows, on the other hand, lead to rapid forgetting and may lead to model overfitting. Active learning approaches such as contrastive learning OOD methods (CADE and HICL), problem-specific sample selection, or uncertainty sampling are particularly useful because they reduce the number of samples that need to be labeled without compromising detection performance. Additionally, they do not include a forgetting mechanism, which helps mitigate the impact of reappearing application behaviors. However, it is worth noting that CADE and HICL involve higher costs since they require generating a new CL model at each retraining step for sample selection, whereas the cost of the problem-specific sample selection method can be considered negligible because clusters are updated incrementally, involving a one-pass process over the data.

As for future work, we propose exploring more complex sliding window mechanisms, such as adapting the size of the sliding window as a function of the distribution dynamics. This mechanism could be useful for dealing with applications that manifest themselves in different ways, e.g., periodically or recurrently. Similarly, more advanced sample selection policies can be explored. These could include, for example, selective forgetting since, on current configurations the set of samples continuously grows.

## CRediT authorship contribution statement

**Borja Molina-Coronado:** Conceptualization, Methodology, Software, Formal analysis, Investigation, Writing - Original Draft, Writing - Review & Editing **Usue Mori:** Conceptualization, Methodology, Writing - Review & Editing **Alexander Mendiburu:** Conceptualization, Methodology, Writing - Review & Editing **Jose Miguel-Alonso:** Conceptualization, Methodology, Writing - Review & Editing

## Acknowledgments

This work has received support from the following programs: PID2019-104966GB-I00AEI (Spanish Ministry of Science and Innovation), IT-1504-22 (Basque Government), KK-2021/00095 and KK-2021/00065 (Elkartek projects SIGZE and ALUSMART supported by the Basque Government). Borja Molina-Coronado holds a predoctoral grant (ref. PRE_2021_2_0230) by the Basque Government.

## A. Analysis of the effect of the parameters

### A.1. Changing the $\lambda$ parameter of the PH method

This section describes the behavior of different parameter configurations for the PH test used for concept drift detection. Specifically, we experiment with different $\lambda$ values of 0.02, 0.03, 0.04 and 0.05. Remember that this parameter determines the maximum allowed decay of the performance of models to raise a retraining signal. For this experiment, we retrain models with all the data available once the retraining signal is flagged. As we can see in Figure 7, there are small differences among the configurations in terms of performance and number of updates. We observe that some lines overlap because the retraining signal and thus, the model, match for different $\lambda$ values. In all cases, a lower $\lambda$ value involves greater sensitivity to changes and thus, a slightly higher number of updates is required. A lower $\lambda$ value is also related with a more stable performance pattern over time. Accordingly, model performance tends to be higher with sensitive (lower) $\lambda$ values. The exception to this rule is MaMaDroid when using $\lambda = 0.05$. In any case, the differences are minimal and results seem robust and coherent for all $\lambda$ values.

### A.2. Changing the $\epsilon$ parameter of the problem-specific sample selection

This section describes the behavior of the models for different parameter configurations of the problem-specific sample selection approach. Specifically, we experiment with different $\epsilon$ radius values for cluster admission of 0, 0.01, 0.02, 0.03, 0.04 and 0.05. Models are retrained regularly, at fixed periods, with the data selected by the sample selection method. As we can see in Figure 8, there are almost no differences among the configurations in terms of performance. If we choose to create smaller clusters, they are conformed of very similar samples and thus, more samples are rejected by the admission condition. This entails the creation of more clusters, involving a larger amount labeling effort but also a reduction of the (automated) labeling errors. Overall, the best trade-off between performance and labeling requirements is obtained with an $\epsilon$ radius of 0.01.

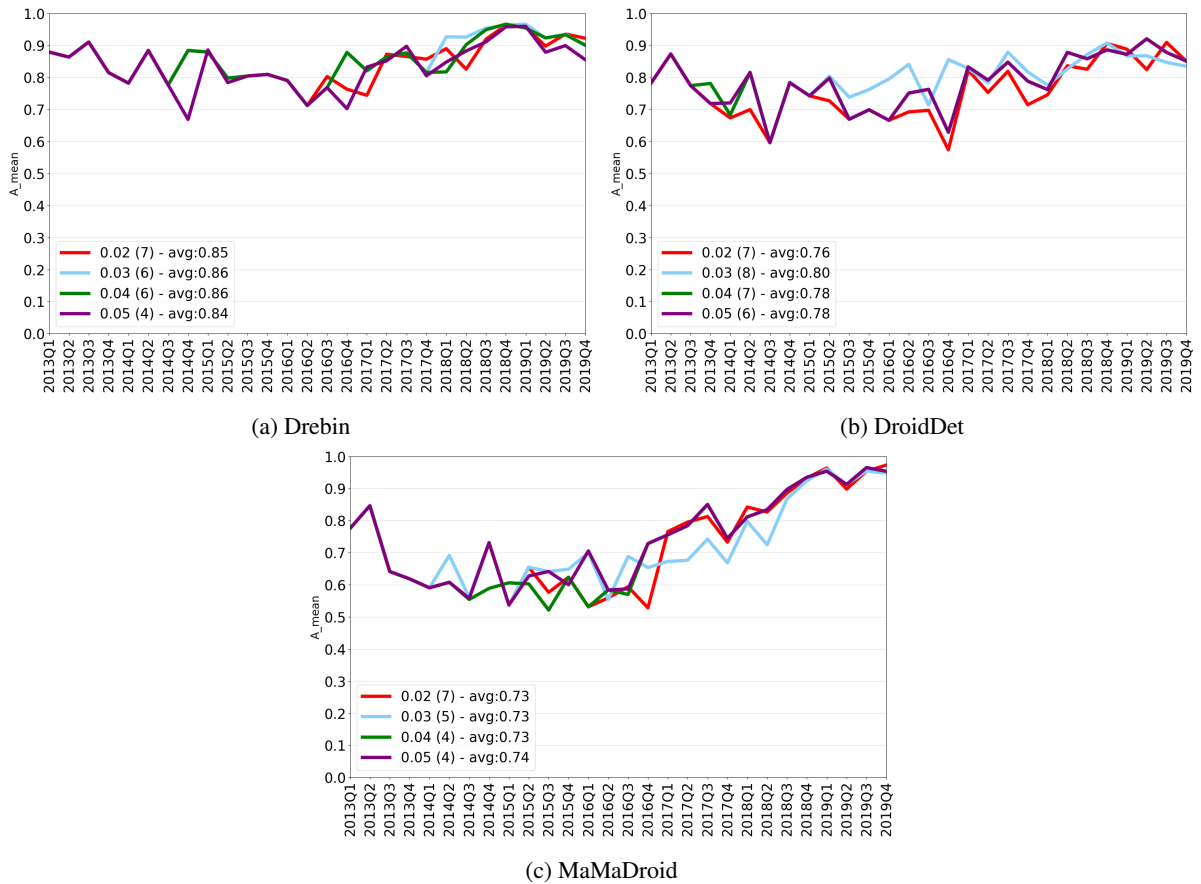

**Figure 7:** Evolution of the performance of detectors with change detection based on the PH test with different $\lambda$ threshold values. Retraining is performed using all the available data. In parentheses, the number of updates required by the Change Detection method.

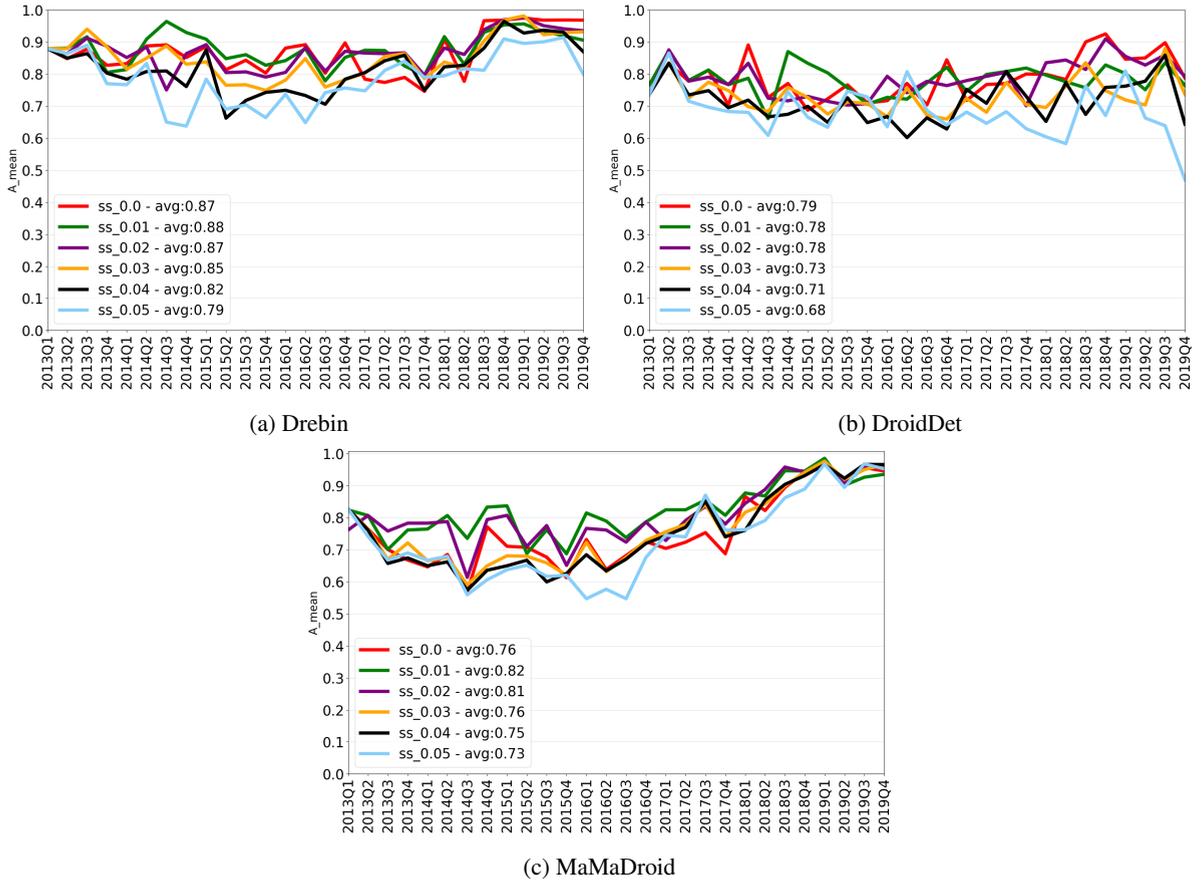

**Figure 8**: Evolution of the performance of detectors with periodical retraining and different problem-specific sample selection radius values ($\epsilon$).

## B. Analysis of the Combined Effect of Change Detection and Sample Selection Methods

When change detection and sample selection policies are combined in retraining (see Table 3), a slight decrease in performance is obtained with respect to retraining at fixed periods (see Table 2) for all sample management policies. Again, as can be seen in the Figure 9, the best approach is the problem-specific method that uses a budget of samples to label of 70%, but other active learning configurations are not far behind. The worst performing approach is the one using the last 100 samples. In terms of the number of model updates, all configurations required a similar number of retraining operations (between 4 and 10) which are far less than using periodic retraining (28). Bearing all this in mind, the conclusion are in the line with those presented in Section 8.

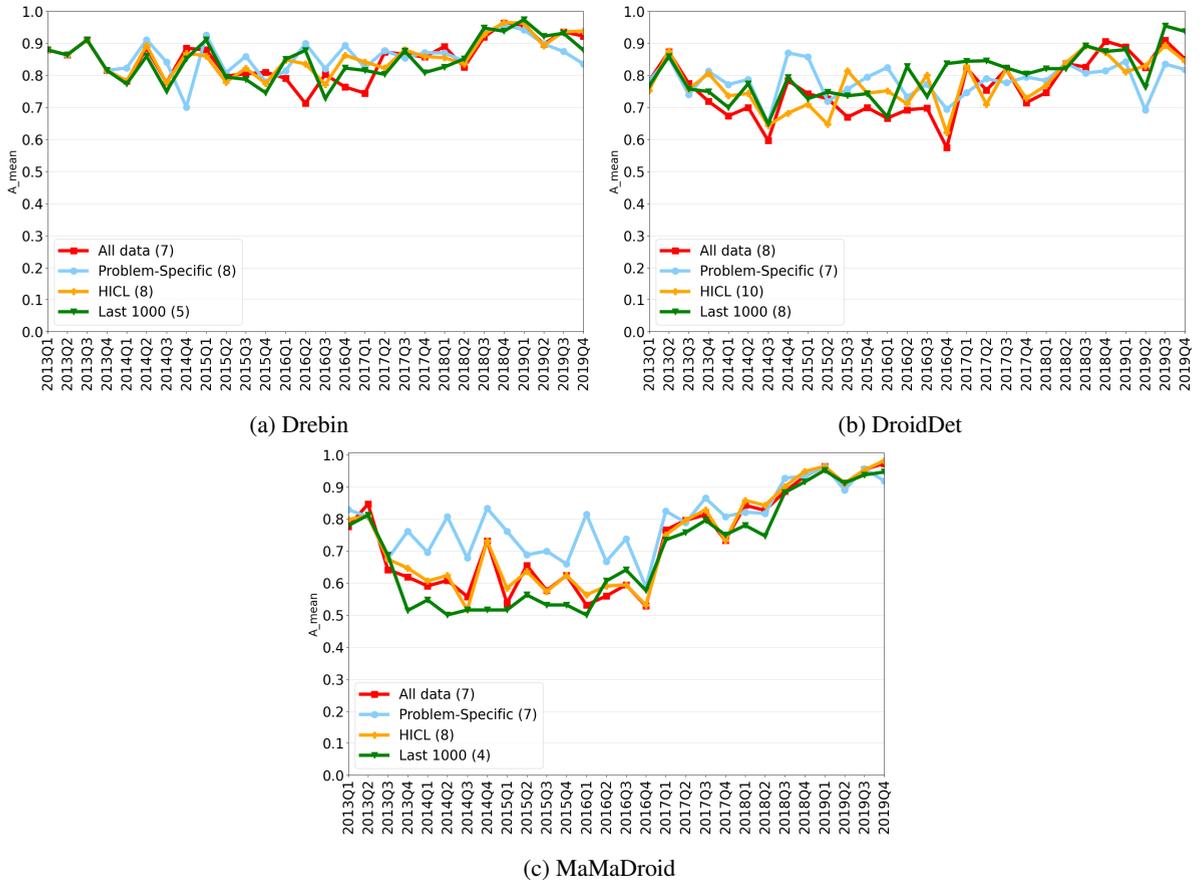

**Figure 9:** Evolution of the performance of detectors with change detection for the period 2013-2019. Numbers between parentheses depict how many retraining and model replacement rounds are triggered by the change detection method to keep the effectiveness of detectors over time.

|  | Drebin | DroidDet | MaMaDroid | Avg. Perf | Updates |
|---|---|---|---|---|---|
| **All data** | 0.85 | 0.76 | 0.72 | 0.77 | 7, 8, 7 |
| **Last 100** | 0.74 | 0.68 | 0.52 | 0.64 | 5, 7, 1 |
| **Last 1000** | 0.85 | 0.80 | 0.69 | 0.78 | 5, 8, 4 |
| **Last 2000** | 0.86 | 0.75 | 0.66 | 0.75 | 8, 8, 4 |
| **Problem-Specific** | 0.86 | 0.78 | 0.79 | 0.81 | 8, 7, 7 |
| **CADE** | 0.79 | 0.74 | 0.72 | 0.75 | 7, 7, 7 |
| **HICL** | 0.86 | 0.77 | 0.73 | 0.78 | 8, 10, 8 |
| **Uncertainty** | 0.84 | 0.77 | 0.72 | 0.77 | 8, 9, 8 |

**Table 3**
Average $A_{mean}$ performance throughout the evaluation period for different sample selection policies and change detection. Column "Updates" represents the number of changes detected by the PH test method for each of the evaluated detectors